\documentclass[final,5p,times,twocolumn]{elsarticle}
\usepackage{amssymb}
\usepackage{amsmath}
\usepackage{color}
\usepackage{kotex} 

\newcommand{\rev}[1]{{\color{black} #1}}

\usepackage[dvipsnames]{xcolor}
\usepackage{ulem}

\usepackage{multirow}

\journal{Journal of Subatomic Particles and Cosmology}

\begin{document}

\begin{frontmatter}

\title{Heavy-ion collision simulation with high performance computer}

\author[label1]{Dae Ik Kim}
\author[label1]{Chang-Hwan Lee}
\author[label2]{Youngman Kim}
\author[label3]{Sangyong Jeon}

\affiliation[label1]{organization={Department of Physics},
             addressline={Pusan National University},
             city={Busan},
             postcode={46241},
             country={Korea}}

\affiliation[label2]{organization={Center for Exotic Nuclear Studies},
             addressline={Institute for Basic Science},
             city={Daejeon},
             postcode={34126},
             country={Korea}}

\affiliation[label3]{organization={Department of Physics},
             addressline={McGill University},
             city={Montreal, Quebec},
             postcode={H3A2T8},
             country={Canada}}

\begin{abstract}
Heavy-ion collision is an important tool to understand the dense nuclear matter properties. 
In order to understand the results of the heavy-ion collision experiments, both theoretical approaches to dense nuclear matter using effective models and the computer simulations with given theoretical models have been performed.
Due to the complexity of the system and the theoretical framework, the heavy-ion collision simulations require heavy computer resources. In this talk, we report our recent preliminary work on the heavy-ion collision simulation using DaeJeon Boltzmann-Uehling-Uhlenbeck (DJBUU) and Sindong Quantum Molecular Dynamics (SQMD) model with high performance computers (HPC).
\end{abstract}




\end{frontmatter}

\section{Motivation} 
\label{sec1}
Detection of gravitational waves GW170817 and electromagnetic afterglows which were generated by the merger of a neutron star binary opened a new era of multi-messenger astrophysics~\cite{LIGOScientific:2017vwq},
and the tidal deformability of neutron stars estimated using the gravitational waves GW170817 demonstrated the potential for  astrophysical observations to understand the properties of dense nuclear matter~\cite{LIGOScientific:2018hze, LIGOScientific:2018cki}.
Additionally, NICER X-ray observations provide the potential to further understand the internal structure of neutron stars by simultaneously estimating their mass and radius~\cite{Miller:2019cac}. These recent developments have made neutron stars  very important astrophysical objects for both astrophysics and nuclear physics.

Properties of dense nuclear matter inside neutron stars can also be estimated by terrestrial heavy-ion collision experiments~\cite{Tsang:2019mlz, Sorensen:2023zkk}. Many heavy-ion collision experiment facilities have been constructed worldwide to test dense matter properties beyond normal nuclear matter density. 
The Rare isotope Accelerator complex for ON-line experiment (RAON) is 
\rev{one such heavy-ion collision facility being built in Korea~\cite{Tshoo:2013voaRAON}. This work mostly concerns heavy-ion collisions that will be conducted at RAON.
}

In order to understand and predict the results of heavy-ion collision experiments, we have performed numerical simulations with DaeJeon Boltzmann-Uehling-Uhlenbeck (DJBUU) code and Sindong Quantum Molecular Dynamics (SQMD) code using high performance 
\rev{computing resources (HPC)} 
provided by Korea Institute of Science and Technology Information (KISTI).

In this presentation, we review our recent results of heavy-ion collision simulations with DJBUU and SQMD using HPC in KISTI. In Sec.~\ref{sec2}, frameworks of heavy-ion collision simulations are briefly introduced. In Sec.~\ref{sec3}, our preliminary numerical results of heavy-ion collision simulations are summarized. In Sec.~\ref{sec4}, we summarize our work and discuss future works.

\section{Framework of heavy-ion collision simulation}
\label{sec2}
Transport models simulate the 
full time-evolution of the
heavy-ion collision dynamics with microscopic degree of freedom, allowing us to study the relationship between the nuclear matter produced in the intermediate stage of the simulation and observables obtained from the final configuration.

One can broadly categorize current transport models into two types, the
Boltzmann-Uehling-Uhlenbeck (BUU) types~\cite{Bertsch1984}, and Quantum Molecular Dynamics (QMD) types~\cite{Aichelin1991}.
In order to study heavy-ion collision experiments at RAON, we have developed two transport models belonging to each category, DaeJeon Boltzmann-Uehling-Uhlenbeck (DJBUU) and Sindong Quantum Molecular Dynamics (SQMD) which are described briefly below.

\subsection{DaeJeon Boltzmann-Uehling-Uhlenbeck}
DJBUU model uses 
\rev{the relativistic BUU equation given by}
\begin{equation}
    \frac{1}{p^{*0}}\big[
    p^\mu\partial_\mu - (p_\mu \mathcal{F}^{\mu i} - m^*\partial^im^*) \frac{\partial}{\partial p^i} 
    \big]f(\vec{x},\vec{p})
    = C(\vec{x},\vec{p})
\label{eq:BUU_Eq}
\end{equation}
where $p^\mu$ is 
\rev{the four momentum of a particle,
$\mathcal{F}^{\mu \nu} = \partial^\mu V^\nu-\partial^\nu V^\mu$ is 
the field strength tensor of the vector mean field(s) $V^\mu$,
$m^*$ is the
effective nucleon mass, and $f(\vec{x},\vec{p})$ is phase-space density of system. The collision term $ C(\vec{x},\vec{p})$ describes baryon-baryon collision processes, elastic and inelastic scatterings. When the collision term is ignored, the above equation corresponds to the Vlasov equation, which is a semi-classical approximation of time-dependent relativistic Hartree-Fock equation.
}

By using test particle method~\cite{Wong1982} widely used for BUU-type models, above BUU equation can be numerically solved. In the test particle method, nucleons are treated as an ensemble of 
\rev{$N_{TP}$ number of test particles.
Each test particle then propagates and/or undergoes collisions according to the mean-fields and the appropriately scaled cross-sections in Eq.(\ref{eq:BUU_Eq}).
}
The phase-space density is obtained as the sum of test particles.
\begin{equation}
    f(\vec{x},\vec{p}) = \frac{(2\pi)^3}{N_{TP}}\sum^{AN_{\rev{TP}}}_i g(\vec{x}-\vec{x}_i)g(\vec{p}-\vec{p}_i)
\end{equation}
where $A$ is the number of nucleons composing projectile and target, $N_{TP}$ is number of test particles per nucleon. 
\rev{Here, $g(\vec{x})$ is a polynomial of the coordinate or momentum profile of test particles, which is a distinctive} 
feature of DJBUU model compared to other BUU-type models 
\rev{that use Gaussians or triangle profiles.
}
The advantage of using polynomial profiles is that, unlike Gaussians, they have finite endpoints and are readily integrable, while providing a smoother shape than triangular profiles.

Propagation in mean-field is described by solving equation of motion of 
\rev{the test particles given by}
\begin{equation}
    \dot{\vec{x}}=\frac{\vec{p}}{p^{*0}}, ~~ \dot{\vec{p}}= -\nabla V^0-\frac{m^*\nabla m^* }{p^{*0}}
\end{equation}
where vector field strength $V^i$ and effective mass $m^*$ are obtained from relativistic mean field (RMF) theory. The Lagrangian density is 
\begin{equation}
    \begin{aligned}
        \mathcal{L} &= \bar{\psi}\big[i\gamma_{\mu}\partial^{\mu} -m_N^* -g_\omega\gamma_\mu\omega^\mu-g_\rho\gamma_\mu\vec{\tau}\cdot\vec{\rho}^\mu-\frac{e}{2}(1+\tau^3)A^\mu\big]\psi \\
        &\quad + \frac{1}{2} (\partial_\mu\sigma \partial^\mu\sigma - m_\sigma^2\sigma) - U_\sigma(\sigma) - \frac{1}{4}\Omega_{\mu\nu}\Omega^{\mu\nu} + \frac{1}{2}m_\omega^2\omega_\mu\omega^\mu \\
        &\quad - \frac{1}{4}R_{\mu\nu}R^{\mu\nu} + \frac{1}{2}m_\rho^2\vec{\rho}_\mu\cdot\vec{\rho}^\mu +\frac{1}{4}F_{\mu\nu}F^{\mu\nu}
    \end{aligned}
\end{equation}
where $\psi$ is the nucleon field, and $\sigma$, $\omega^\mu$, and $\vec\rho^\mu$ are meson field of 
\rev{the scalar-isoscalar meson $\sigma$, vector-isoscalar meson $\omega$, and vector-isovector meson $\rho$, respectively. The electromagnetic field is represented by $A^\mu$.}
Additionally, $g_\sigma$, $g_\omega$, and $g_\rho$ are nucleon-meson couplings, and 
\rev{the field strength tensors are}
\begin{align}
    \Omega^{\mu\nu}=\partial^{\mu}\omega^\nu-\partial^\nu\omega^\mu \\
    R^{\mu\nu}=\partial^{\mu}\rho^\nu-\partial^\nu\rho^\mu\\
    F^{\mu\nu}=\partial^{\mu}A^\nu-\partial^\nu A^\mu.
\end{align}
$U(\sigma)$ is the non-linear potential energy of the $\sigma$ field. 
\rev{The original formulation of DJBUU used quantum hadrodynamics (QHD) model \cite{Liu:2001iz} in which 
the $\sigma$ potential and the effective nucleon mass are given by}
\begin{equation}
    m_N^* = m_N - g_\sigma \sigma, \quad U(\sigma)=\frac{1}{3}g_2\sigma^3+\frac{1}{4}g_3\sigma^4.
\end{equation}
\rev{In the QHD model, \( U(\sigma) \) is usually introduced \((U(\sigma) \neq 0)\) because, without it, incompressibility \( K_0 \) becomes too large (typically \( K_0 > 500~\mathrm{MeV} \)). The inclusion of \( U(\sigma) \) allows the model to reproduce a more reasonable value of \( K_0 \) in the range of \( 200\text{--}300~\mathrm{MeV} \).
}

In this work, we alternatively adopt the Quark-Meson Coupling (QMC) model in which additional effects of the Bag constant are taken into account. In this model, the nucleon effective mass is given by
\begin{equation}
    m_N^* = m_N - g_\sigma \sigma\bigg(1-\frac{a_N}{2}g_\sigma \sigma\bigg).
\end{equation}
$a_N$ is obtained from a quark-level calculation of the bag energy in external meson fields, which determine the nucleon's effective mass. 
The bag energy is given by
\begin{equation}
m_N^{*} = 
\sum_{q} \frac{n_q \Omega_q^{*} - z_0}{R^{*}} 
+ \frac{4}{3} \pi (R^{*})^3 B,
\label{eq:bag_energy}
\end{equation}
subject to the stability condition ${d m_N^{*}}/{d R^{*}} = 0$.
$B$ is the bag constant, $z_0$ accounts for center-of-mass and gluon fluctuation corrections. $n_q$ is the number of quarks in Bag, and 
\begin{equation}
\Omega_q^{*} = \sqrt{ (x_q^{*})^2 + (m_q^{*} R^{*})^2 }, 
\quad
m_q^{*} = m_q - g^q_\sigma \sigma,
\end{equation}
where $m_q^{*}$ is the effective quark mass under the scalar mean field $\sigma$, and $x^*_q$ is obtained from boundary condition of the bag surface $j_0(x^*_q)=\beta^{*}_qj_1(x^*_q)$. $j_{0}$ and $j_1$ are spherical Bessel function and $\beta^*_q=\sqrt{{(\Omega^*_q-m^*R^*})/({\Omega^*_q+m^*R^*})}$. 
Additional quadratic term of $\sigma$ in the effective mass in the QMC model plays a role similar to the self-interaction potential \( U(\sigma) \) in QHD. Therefore, the QMC model can reproduce a sufficiently low incompressibility $K_0$ even without introducing $U(\sigma)$.
We use parameters including $m_\sigma$, coupling constants, and $a_N$ for the QMC model from Ref.~\cite{Tsushima:2022PTEP}.

Collisions between nucleons (test particles) are described as semi-classical collisions. They are detected using a common criterion called Bertsch's prescription~\cite{Bertsch:1988ik} and some of them are declined following the Pauli blocking factor.
The cross-section for elastic collisions and  inelastic collisions are provided in Refs.~\cite{Cugnon:1996kh, Bertsch:1988ik} respectively.
\rev{Further details of the DJBUU model can be found in Ref.~\cite{Kim:2020sjy}.}

\subsection{Sindong Quantum Molecular Dynamics}
The SQMD model also simulates heavy-ion collision dynamics with 
propagations and collisions just like DJBUU, but there are 
\rev{some important differences between two models.}
SQMD utilizes the QMD approach allowing the $n$-body dynamics, 
\rev{where as the BUU models deal with the dynamics of the 1-body distribution.}
Nucleons are treated as Gaussian wave packet with a fixed width. 
The phase-space density is derived from the Wigner transformation of these wave packets.

We use density-dependent Skyrme potential for nuclear interaction between nucleons.
\begin{equation}
    U_{\mathrm{Skyrme}} = \frac{\alpha}{2} \bigg(\frac{\rho}{\rho_0} \bigg) 
    + \frac{\beta}{\gamma+1} \bigg(\frac{\rho}{\rho_0} \bigg)^\gamma
\end{equation} 
$\alpha$, $\beta$, $\gamma$ are respectively $-218$ MeV, 164 MeV and $4/3$ and $\rho$ is the baryon density and $\rho_0$ is the saturation density of 0.16 fm$^{-3}$.

The classical Hamiltonian is
\begin{equation}
    H = \sum_i\frac{\vec{p_i}^2}{2m_i} + U_{\mathrm{Skyrme}}+U_{\mathrm{surf}}+U_{\mathrm{sym}}+U_{\mathrm{Coulomb}}
\end{equation}
where the first term is sum of the kinetic energy of $i$th nucleon with mass $m_i$ and momentum $\vec{p_i}$ and $U_{\mathrm{surf}}$ and $U_{\mathrm{sym}}$ are surface and symmetry terms.

\rev{At the propagation stage}, 
the nucleons follow equation of motion derived from the Hamiltonian.
\begin{equation}
\frac{d \vec{r}}{d t}=\nabla_{\vec{p}} H , \quad \frac{d \vec{p}}{d t}=-\nabla_{\vec{r}} H,
\end{equation}
where $\vec{r}$ and $\vec{p}$ are the position and momentum vector of the center of nucleons, respectively.

\rev{The realization of collisions in SQMD model is almost the same as that of the DJBUU model. In the present version of SQMD, we consider only the elastic collisions between the nucleon with the $NN$ scattering cross-section from Ref.~\cite{Li:1993ef,Li:1993rwa}. In future versions, inelastic cross-section will be added to study the near- or sub-threshold pion productions.}

\rev{For the fragmentation and clustering, we use the commonly used Minimal Spanning Tree algorithm. Further details of the SQMD model can be found in Ref.~\cite{Kim:2017hmt}.}

\section{Numerical results}
\label{sec3}

Heavy-ion collision simulations using transport models require huge computing resources 
\rev{due to the large number of particles, short time steps, and need to accumulate a statistically meaningful number of events.}
For example, in a typical DJBUU simulation of the $^{196}$Au+$^{196}$Au collision lasting 140 fm/$c$ with a time interval of 0.2 fm/$c$ and 100 test-particles per nucleon in a 100 fm $\times$ 100 fm $\times$ 100 fm cubic box, the total number of grid cells, particles and time steps are 1,000,000, 39,200 and 700, respectively. 

At each time step, we consider the possibility of hard collisions among these particles, 
\rev{and in every grid cell, 
we calculate the baryon density, scalar density, and meson mean fields from the overall particle distribution.} 

In the case of SQMD simulations, the number of particles is smaller than 
\rev{that} in DJBUU, but far more events are used 
\rev{for meaningful statistics.}
In order to handle these computationally expensive simulations, we use the HPC system NURION provided by KISTI. Below are results including preliminary results obtained using NURION.

\subsection{A comparative study of DJBUU and SQMD} 

\begin{table*}[ht]
\renewcommand{\arraystretch}{1.4}
\centering
\caption{Comparison of BFs produced in $^{208}$Pb + $^{40,48}$Ca simulations: the five most abundant BFs from DJBUU runs and the three most abundant BFs from SQMD runs.}
\label{tab:PbCaFragment}
\begin{tabular}{c|c|c|c|c}
\noalign{\smallskip}\noalign{\smallskip}
\hline
$~~$Target$~~$ & \multicolumn{1}{c|}{\begin{tabular}[c]{@{}c@{}}$E_{\rm beam}$ {[}AMeV{]}\end{tabular}} & \multicolumn{1}{c|}{\begin{tabular}[c]{@{}c@{}} $b$  {[}fm{]}\end{tabular}}& DJBUU  &  SQMD \\ 
\hline
\multirow{6}{*}{$^{40}$Ca} & \multirow{3}{*}{50} & 0 &$^{163}_{73}$Ta,\;$^{162}_{73}$Ta,\;$^{164}_{73}$Ta,\;$^{163}_{74}$W & $^{163}_{69}$Tm,\;$^{173}_{74}$W,\;$^{169}_{72}$Hf\\ 
      & & 3 &$^{163}_{73}$Ta,\;$^{165}_{74}$W,\;$^{164}_{73}$Ta,& $^{169}_{72}$Hf,\;$^{173}_{74}$W,\;$^{172}_{74}$W\\ 
      & & 6 &$^{167}_{74}$W,\;$^{169}_{75}$Re,\;$^{165}_{73}$Ta,\;$^{168}_{75}$Re & $^{168}_{72}$Hf,\;$^{164}_{70}$Yb,\;$^{169}_{72}$Hf\\ 
      \cline{2-5}
   & \multirow{3}{*}{100} & 0 &$~~~$ $^{123}_{56}$Ba,\;$^{121}_{55}$Cs,\;$^{124}_{57}$La,\;$^{122}_{56}$Ba,\;$^{124}_{56}$Ba $~~~$ & $~~~$ $^{78}_{33}$As,\;$^{114}_{50}$Sn,\;$^{124}_{54}$Xe $~~~$\\ 
      & & 3 &$^{130}_{59}$Pr,\;$^{130}_{58}$Ce,\;$^{128}_{57}$La,\;$^{128}_{58}$Ce,\;$^{129}_{58}$Ce
      & $^{125}_{53}$I,\;$^{128}_{56}$Ba,\;$^{132}_{57}$La\\ 
      & & 6 &$^{145}_{64}$Gd,$^{144}_{64}$Gd,$^{146}_{65}$Tb,$^{147}_{65}$Tb & $^{151}_{64}$Gd,$^{149}_{63}$Eu,$^{154}_{66}$Dy\\
      \cline{1-5}
\multirow{6}{*}{$^{48}$Ca} & \multirow{3}{*}{50} & 0 &$^{161}_{72}$Hf,$^{162}_{72}$Hf,$^{160}_{71}$Lu,$^{159}_{71}$Lu & $^{167}_{70}$Yb,$^{167}_{71}$Lu,$^{170}_{71}$Lu\\
      & & 3 &$^{162}_{72}$Hf,$^{164}_{73}$Ta & $^{165}_{70}$Yb,$^{167}_{70}$Yb,$^{167}_{71}$Lu\\
      & & 6 &$^{164}_{72}$Hf,$^{163}_{72}$Hf,$^{166}_{73}$Ta,$^{165}_{72}$Hf & $^{165}_{69}$Tm,$^{159}_{68}$Er,$^{164}_{69}$Tm\\
      \cline{2-5}			
  & \multirow{3}{*}{100} & 0 &$^{113}_{51}$Sb,$^{115}_{52}$Te,$^{114}_{51}$Sb,$^{116}_{52}$Te,$^{112}_{51}$Sb & $^{58}_{25}$Mn,$^{74}_{32}$Ge,$^{107}_{48}$Pd\\
      & & 3 &$^{121}_{54}$Xe,$^{122}_{55}$Cs,$^{120}_{54}$Xe,$^{123}_{55}$Cs,$^{121}_{55}$Cs & $^{120}_{52}$Te,$^{106}_{45}$Rh,$^{113}_{48}$Cd\\
      & & 6 &$^{140}_{62}$Sm,$^{139}_{62}$Sm,$^{138}_{61}$Pm,$^{137}_{61}$Pm,$^{137}_{60}$Nd & $^{147}_{62}$Sm,$^{153}_{64}$Gd,$^{148}_{62}$Sm\\
      \hline
\end{tabular}
\end{table*}

In this study, we compare results from two transport models, DJBUU and SQMD, focusing on the production of the biggest fragments (BFs) in heavy-ion collisions between stable nuclei. To this end, we simulated $^{208}$Pb + $^{40,48}$Ca at $E_{\rm beam} =$ 50 and 100$~\hbox{AMeV}$, energies that lie within the overlap region where both models are applicable. Notably, we use different definitions of the BF, since clustering is not implemented in DJBUU as it is in SQMD. In DJBUU, the BF is defined as a collection of nucleons satisfying the density criterion $\rho_B/\rho_0 > 0.1$ at the final configuration, whereas in SQMD it is defined as the fragment with the largest mass number among those produced.

Table \ref{tab:PbCaFragment} shows that the two models produce similarly sized BFs at 50 AMeV, 
\rev{but the model dependence is relatively larger at the higher beam energy (100 AMeV) and smaller impact parameters since
the equations of state, the definitions of the BF, and the intrinsic difference between the BUU-type and the QCD type models are more pronounced at the higher energy \cite{Kim:2022zdn}.}

Additionally, to study heavy-ion collisions with an isotope beam that can be produced at the RAON accelerator, we performed $^{20}$Na + $^{208}$Pb simulations and found that the model dependence becomes more pronounced with an unstable projectile.
\rev{Result of DJBUU simulation show about 30\% bigger BFs than one of SQMD simulation \cite{Kim:2024eti}.  One way to understand such model differences is to examine the stability of the projectile as simulated by each model. To examine the stability of the density profile of the unstable nucleus \(^{20}\mathrm{Na}\), we performed simulations using a very large impact parameter of 20 fm so that the nuclei propagate without actual collision in both DJBUU and SQMD models. Initially, both models showed a similar density profile, but after about 20 fm/\(c\), DJBUU exhibited a tendency for the central density to become more compact, while SQMD showed the opposite tendency, with the central density decreasing and spreading out. This behavior suggests that the relatively low stability of the unstable \(^{20}\mathrm{Na}\) leads to rapid changes in the density profile, and the opposite evolution trends in the two models contributed to the observed differences in the results.}

These results allow us to discuss how our transport models, DJBUU and SQMD, are further developed to simulate rare isotope beam heavy-ion collisions \cite{Kim:2024eti}.

\subsection{Adopting the QMC model in DJBUU}
Since P. Guichon developed the Quark-Meson Coupling (QMC) model as the extension of Quantum Hadron Dynamics (QHD) \cite{Guichon1988}, the QMC model has been implemented in various field, such as finite nuclei, nuclear matter and neutron star \cite{Guichon2018, Stone2016, Stone2017, Pal1999}. In this work, we study the QMC model in heavy-ion collisions by adopting it in the DJBUU transport model. For clarity, we refer to the original DJBUU model, which implemented the QHD model, as DJBUU+QHD, and to the DJBUU variant, which recently adopted the QMC model, as DJBUU+QMC.

To benchmark the validity of the QMC model, we simulate the same system using both models and compare the results. 
We simulate $^{40}$Ca + $^{40}$Ca system at $E_\mathrm{beam}$ = 100 AMeV with impact parameter $b$ = 0 fm.

\begin{figure}[h]
\centering
\vspace{1em}
\includegraphics[width=0.95\linewidth]{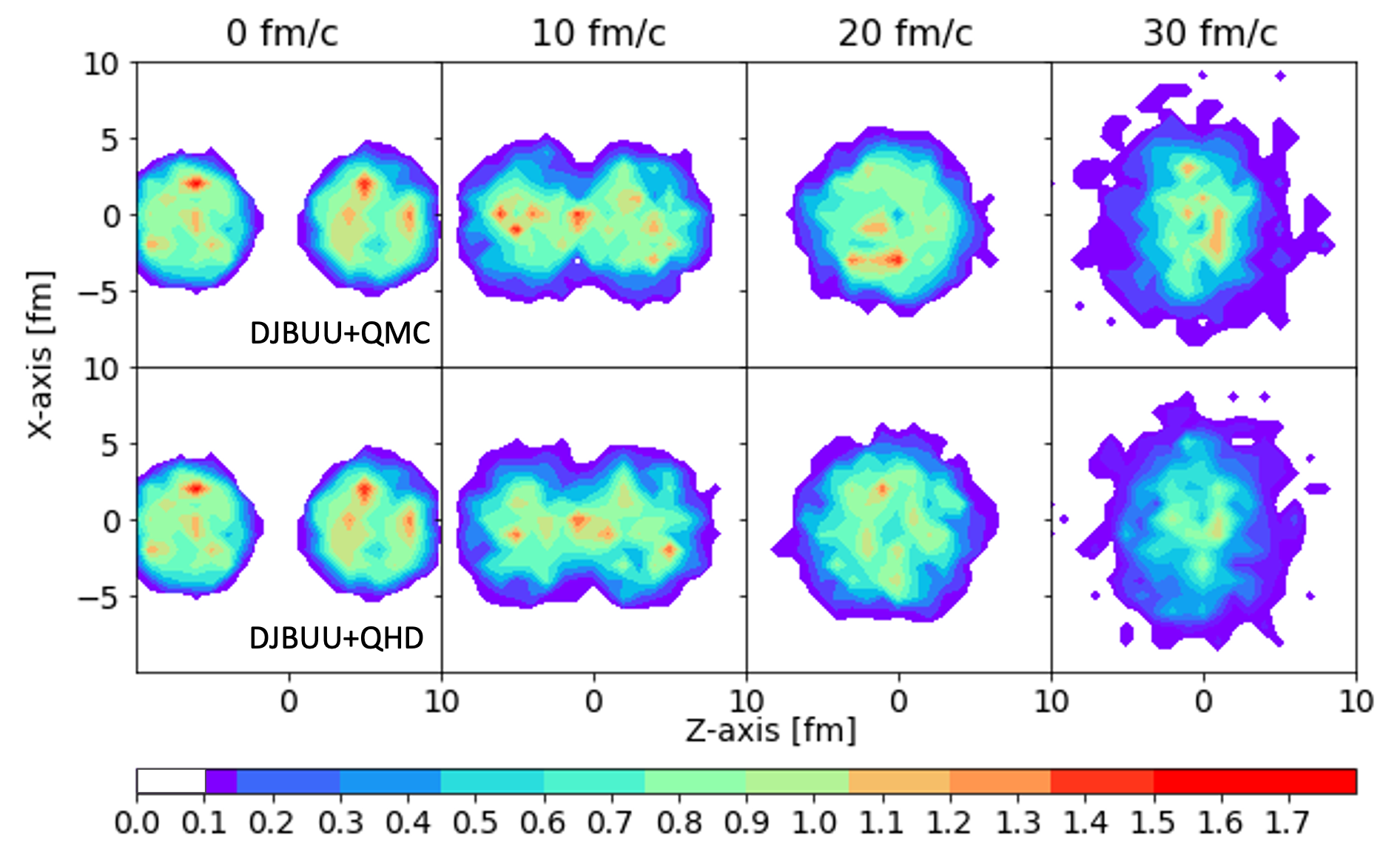}
\caption{Time evolution of density contours in the reaction plane for $^{40}$Ca + $^{40}$Ca collisions at an impact parameter of $b = 0$ fm and a beam energy of 100 AMeV. The upper and lower panels show results from DJBUU+QHD and DJBUU+QMC, respectively.}
\label{fig:Contour}
\vspace{1em}
\end{figure}

\begin{figure}[h]
\centering
\vspace{1em}
\includegraphics[width=0.95\linewidth]{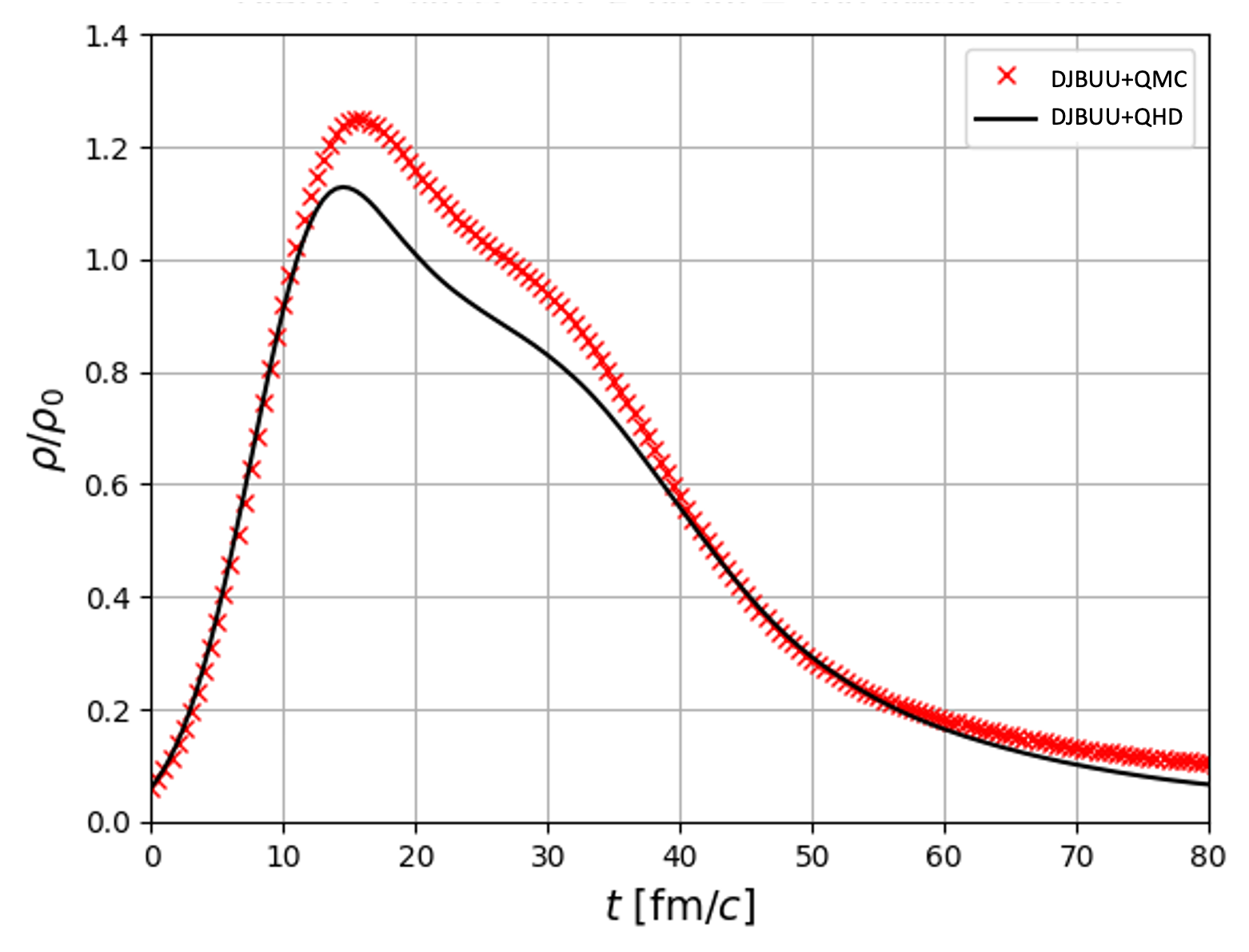}
\caption{Time evolution of baryon density at the origin of the center-of-mass frame for $^{40}$Ca + $^{40}$Ca collisions at an impact parameter of $b = 0$ fm and a beam energy of 100 AMeV. The black solid line and red crosses correspond to the results from DJBUU+QHD and DJBUU+QMC, respectively.}
\label{fig:density}
\vspace{1em}
\end{figure}

Figures \ref{fig:Contour} and \ref{fig:density} show results from DJBUU+QMC similar to those from DJBUU+QHD. 
\rev{The similarities between two results
suggest that the QMC model is correctly implemented. 
To show the difference, we plot the maximum nucleon density calculated in the two models in Fig.\ref{fig:density}. It is seen that the maximum density in DJBUU+QMC is higher than that in DJBUU+QHD.}
\rev{This difference could lead to the differences in the pion multiplicity
which is known to be sensitive to the equation of state of nuclear matter. Calculations and analysis of the pion production are in progress.}

\section{Discussions}
\label{sec4}
In this work, we have shown that heavy-ion collision simulations with the DJBUU and SQMD transport models on HPC systems provide valuable opportunities to study dense nuclear matter, as expected to be produced at the RAON experiments.
\rev{In case of $^{208}$Pb+${^{40,48}}$Ca simulation,} our comparative study indicates that while both models produce consistent results at the lower beam energy, there are relatively large differences between results of both models at the higher beam energy. These discrepancies are due to differences in the equation of state and the dynamics of the models.
\rev{$^{208}\mathrm{Pb}+ {}^{40,48}\mathrm{Ca}$ collisions at a beam energy of 50 A MeV show almost the same size of the BF from both models, whereas in the \({}^{20}\mathrm{Na}+{}^{208}\mathrm{Pb}\) case, the BF mass number differed by approximately 30\%. This significant discrepancy appears to originate from differences in how \({}^{20}\mathrm{Na}\) is initialized in each model and how it deforms prior to the collision.} 
\rev{As planned, if the meson field equation solver including surface terms and the time-dependent wave packet width are implemented in DJBUU and SQMD, respectively, the stability of nuclei in simulation is expected to improve, making it possible to describe heavy-ion collisions using unstable nuclei more reliably.}

\rev{Interestingly, recent implementation of the QMC model in the DJBUU model results in a higher maximum density at the collision center compared to the original implementation that used the QHD model.} These results may affect the observables sensitive to the nuclear matter properties, such as direct flow or pion yield ratio.

These studies provide direction for further model refinement to better simulate heavy-ion collisions at rare isotope beam facilities such as RAON. Additionally, by enabling the investigation of the equation of state, which has already been applied to neutron stars and heavy-ion collisions, these studies will contribute to expanding our understanding of dense nuclear matter alongside gravitational wave observations.

\section*{Acknowledgments}
DIK and CHL were supported by the National Research Foundation of Korea (NRF) grants funded by the Korea government (No. RS-2023-NR076639).
SJ acknowledges the support of the Natural Sciences and Engineering Research Council of Canada (NSERC) [SAPIN-2024-00026].
DIK was supported by the Hyundai Motor Chung Mong-Koo Foundation.

\bibliographystyle{unsrt}
\bibliography{biblio}

\end{document}